\documentstyle[12pt]{article}
\begin{document}

\begin{center}
{\Large\bf SPIN OF CHERN-SIMONS VORTICES}\\[2cm]
{\bf R. Banerjee}\footnote{narayan@bose.ernet.in}\\
{\normalsize S.N.Bose National Centre for Basic Sciences}\\
{\normalsize Block JD, Sector III, Salt Lake City,
Calcutta 700091, India}\\
{\normalsize and}\\
{\bf P. Mukherjee}\\
{\normalsize A.B.N. Seal College}\\
{\normalsize Cooch Behar, West Bengal}\\
{\normalsize India}\\
\end{center}
\vspace{1cm}

\begin{abstract}
We discuss a novel method of obtaining the fractional spin of abelian
and nonabelian Chern-Simons vortices. This spin is interpreted as the
difference between the angular momentum obtained by modifying
Schwinger's energy momentum tensor by the Gauss constraint, and the
canonical (Noether) angular momentum. It is found to be a boundary
term depending only on the gauge field and, hence, is independent of
the matter sector to which the Chern-Simons term couples. Addition of
the Maxwell term does not alter the fractional spin.
\end{abstract}
\newpage
It has been known for quite some time that the introduction of the
Chern-Simons term can lead to vortex solutions in different 2+1
dimensional theories [1-8]. Examples are the occurrence of topological
solitons in O(3) nonlinear sigma model [1-2], vortices in abelian [3]
and nonabelian [4] Higgs models including their generalisation [7]. In
the conventional [3,4] Higgs models, the kinetic term of the gauge
field is given by including both the Maxwell and Chern-Simons terms.
Subsequently, self-dual vortices were also demonstrated in Higgs
models even in the absence of the Maxwell term [5,6,7]. A particularly
interesting feature of these Chern-Simons vortices, which make them a
viable candidate  for anyonlike objects, is that these carry
fractional spin.

In this paper we discuss a novel field-theoretic method of computing
the fractional spin which highlights the  important role played by the
constraints in the theory. To put our analysis in proper perspective,
let us first recall the standard methods [1,2,5,6,8] of computing the
fractional spin. The first is the action oriented approach [1,8]. If
the values of the action for a static and adibatically rotated
configuration are denoted by $S_{st}$ and $S_{ad}$, respectively, then
the spin of the soliton is given by,
$$K = \frac{S_{ad} - S_{st}}{2\pi}$$
The second approach, which has been exemplified in [3-6], is to
calculate the angular momentum in a static configuration. If a nonzero
value is found, one interprets this extra angular momentum as the spin
of the soliton. Both these methods require the knowledge of detailed
(matter and gauge) field configurations obtained by solving the
classical equations of motion and exploiting some viable ansatz.
Furthermore, in models [4,5] involving the Maxwell piece, these
solutions are quite involved and the computation of the fractional
spin becomes less illuminating. Moreover, it is not ascertained
whether the result is insensitive to different choices of ansatz.
Interestingly, however, the expression for the fractional spin in
different models [1-6] turns out to be structurally identical, a fact
which is as remarkable as it is ill understood.

The approach discussed in this paper bypasses the problematic issues
and also illuminates the reason for the structural similarity of the
fractional spin in different theories. Detailed field configurations
are not required. We also do away with the necessity of choosing some
ansatz. The basic idea is to start from Schwinger's [9] definition of
angular momentum which, by construction, is gauge invariant and hence
physically relevant. Next, the canonical part is abstracted from this
angular momentum. This is done by subtracting the Noether (canonical)
angular momentum from the Schwinger angular momentum. The difference,
which is found to be a boundary term, is interpreted as the fractional
spin of the vortex. Moreover, this boundary term is found to be model
independent. Explicit computation of the boundary term, which just
requires the asymptotic form of the gauge field in a rotationally
symmetric configuration, yields the desired fractional spin. A
comparision of our approach with the usual approaches of obtaining
fractional spin has also been done. The extension of our analysis to
nonabelian vortex configurations has been discussed.

Let us first consider the Chern-Simons-Higgs Lagrangian considered in
[5,6],
\begin{equation}
{\cal L} = (D_{\mu}\phi)^{*} (D^{\mu}\phi) + \frac{k}{2}
\epsilon^{\mu\nu\lambda} A_{\mu}\partial_{\nu}A_{\lambda} -
V(\vert\phi\vert)
\end{equation}
where the covariant derivative is defined by,
\begin{equation}
D_{\mu} = \partial_{\mu} + ieA_{\mu}
\end{equation}
and the potential $V(\vert\phi\vert)$ has a symmetry breaking minimum
at $\vert\phi\vert = \vartheta$ and contains only renormalisable
interactions [6]. Note that $A_{\circ}$ is a lagrange multiplier which
enforces the Gauss constraint,
\begin{equation}
G = k\epsilon^{ij}\partial_{i}A_{j} + ie(\phi\pi - \phi^{*}\pi^{*})
\approx 0
\end{equation}
where $\pi(\pi^{*})$ is the momentum conjugate to $\phi(\phi^{*})$,
\begin{equation}
\pi = \frac{\partial {\cal L}}{\partial\dot{\phi}} = \dot{\phi}^{*} -
ieA_{\circ}\phi^{*}
\end{equation}
\begin{equation}
\pi^{*} = \frac{\partial {\cal L}}{\partial\dot{\phi}^{*}} = \dot{\phi} +
ieA_{\circ}\phi
\end{equation}

The Lagrangian (1) is already first order in time derivatives so that
the basic (equal time) brackets can be immediately read-off from
symplectic arguments [10],
\begin{equation}
\{\phi(\vec{x}), \pi(\vec{y})\} = \{\phi^{*}(\vec{x}), \pi^{*}(\vec{y})\} = \delta(\vec{x} -
\vec{y})
\end{equation}
\begin{equation}
\{A_{i}(\vec{x}), A_{j}(\vec{y})\} = \frac{1}{k}\epsilon_{ij}\delta(\vec{x} -
\vec{y})
\end{equation}

The energy-momentum tensor following from Schwinger's [9] definition
is given by,
\begin{eqnarray}
\Theta_{\mu\nu} &= &\frac{\delta S}{\delta g^{\mu\nu}} \nonumber \\
              &= &(D_{\mu}\phi)(D_{\nu}\phi)^{*} +
              (D_{\mu}\phi)^{*}(D_{\nu}\phi) -
              g_{\mu\nu}[(D_{\sigma}\phi)^{*}(D^{\sigma}\phi)-V(\vert\phi\vert)]
\end{eqnarray}
where the Chern-Simons term, being covariant without reference to the
metric, does not contribute. In the presence of the constraint (3) a
more general expression for $\Theta_{\mu\nu}$ follows. This is called
the total energy-momentum tensor [11],
\begin{equation}
\Theta^{T}_{\mu\nu} = \Theta_{\mu\nu} + \wedge_{\mu\nu}G
\end{equation}
where $\wedge_{\mu\nu}$ are multipliers that can be fixed by requiring
that the fields transform normally under the various space-time
generators [12,13]. Note that since G is the generator of
time-independent gauge transformations, the gauge invariance of
$\Theta^{T}_{\mu\nu}$ is preserved on the constraint surface, i.e.,
\begin{equation}
\{\Theta^{T}_{\mu\nu}, G\} \approx 0
\end{equation}
Since we are concerned with the angular momentum operator defined by,
\begin{equation}
J = \int d^{2}x \epsilon^{ij}x_{i} \Theta^{T}_{\circ j}
\end{equation}
let us concentrate on the $(\circ j)$ component of (9). It is easy to
verify, using the algebra (6,7) and the relations (8), (9), that the
correct transformations under spatial translations,
\begin{equation}
\{\chi, \int\Theta^{T}_{\circ i}\} = \partial_{i}\chi
\end{equation}
where $\chi$ generically denotes the basic fields $(\phi, \phi^{*},
\pi, \pi^{*}, A_{i})$, are obtained for the choice,
\begin{equation}
\wedge_{\circ i} = - A_{i}
\end{equation}
One can check that with $\wedge_{\circ\mu} = -A_{\mu}$ the correct
transformation properties follow under all the space time generators
(translations, rotations, boosts) defined from (9).
Hence the desired structure of $J$ simplifies to,
\begin{equation}
J = \int d^{2}\vec{x} \epsilon^{ij}x_{i}[\pi\partial_{j}\phi +
\pi^{*}\partial_{j}\phi^{*} - k\epsilon^{lm}A_{j}\partial_{l}A_{m}]
\end{equation}

We now focus on the canonical angular momentum which requires
Noether's definition of the energy momentum tensor,
\begin{eqnarray}
\theta^{N}_{\mu\nu} &= &\frac{\partial
{\cal L}}{\partial(\partial^{\mu}\phi)} \partial_{\nu}\phi + \frac{\partial
{\cal L}}{\partial(\partial^{\mu}\phi^{*})} \partial_{\nu}\phi^{*} +
\frac{\partial {\cal L}}{\partial(\partial^{\mu}A^{\sigma})}
\partial_{\nu}A^{\sigma} - g_{\mu\nu}{\cal L} \nonumber \\
&= &(D_{\mu}\phi)^{*} \partial_{\nu}\phi +
(D_{\mu}\phi)\partial_{\nu}\phi^{*} +
\frac{k}{2}\epsilon_{\rho\mu\sigma}A^{\rho}\partial_{\nu}A^{\sigma} -
g_{\mu\nu}{\cal L}
\end{eqnarray}
Note that, in contrast to the case of Schwinger's definition (8), the
Chern-Simons term does contribute to $\Theta^{N}_{\mu\nu}$. Once again
it is possible to define, in analogy with (9), a total (canonical)
energy momentum tensor by addition to (15) a term proportional to the
Gauss constraint. This time, however, such a term is absent because
the correct transformations are already generated by (15). Hence the
canonical (Noether) angular momentum can be directly obtained from
(15),
\begin{equation}
J^{N}_{ij} = \int d^{2}\vec{x}(x_{i}\theta^{N}_{\circ j} -
x_{j}\theta^{N}_{\circ i} + \frac{\partial
{\cal L}}{\partial(\partial^{\circ}A^{k})} (g^{k}_{i}g^{l}_{j} -
g^{k}_{j}g^{l}_{i})A_{l})
\end{equation}
which can be simplified in terms of the single component of
$J^{N}_{ij}$ as,
\begin{eqnarray}
J^{N} &= &\int d^{2}\vec{x}\epsilon^{ij}(x_{i}\theta^{N}_{\circ j} +
\frac{k}{2}\epsilon_{il}A^{l}A_{j}) \nonumber \\
      &= &\int d^{2}\vec{x}[\epsilon^{ij}x_{i}(\pi\partial_{j}\phi +
      \pi^{*}\partial_{j}\phi^{*} -
      \frac{k}{2}\epsilon_{lm}A^{l}\partial_{j}A^{m}) +
      \frac{k}{2}A^{j}A_{j}]
\end{eqnarray}
where we have used (15) for simplifications.

Let us next compute the difference between the two angular momentum
operators (14) and (17),
\begin{equation}
K = J - J^{N} = - \frac{k}{2}\int d^{2}\vec{x} \partial^{i}[x_{i}A_{j}A^{j}
- A_{i}x_{j}A^{j}]
\end{equation}
which can be further expressed in a compact form,
\begin{equation}
K = - \int d^{2}\vec{x} \partial^{i}[\pi_{i} \epsilon^{jk} x_{j} A_{k}]
\end{equation}
where
\begin{equation}
\pi_{i} = \frac{\partial {\cal L}}{\partial \dot{A}^{i}} = \frac{k}{2}
\epsilon_{ij} A^{j}
\end{equation}
is the momentum conjugate to $A^{i}$. The above relation is
effectively a constraint which has been accounted by the symplectic
bracket (7). It is now observed that the difference $K$ between the
two expressions of angular momentum is a boundary term. For
nonsingular configurations this vanishes. For singular configurations,
however, $K$ need not vanish. This is precisely what happens for the
Chern-Simons vortices. Indeed the asymptotic form of the gauge
potential, which is all that is required to evaluate (18) or (19), in
a rotationally symmetric configuration is well known [1,14,6],
\begin{equation}
eA_{i} = - \epsilon_{ij} \frac{x^{j}}{\vert x \vert^{2}}n
\end{equation}
where $n$ is a topological invariant called the vortex number which
takes only integer values and is related to the magnetic flux $\Phi$,
\begin{equation}
\Phi = \int d^{2}\vec{x} \epsilon_{ij}\partial_{i}A_{j} = \frac{2\pi n}{e}
\end{equation}
We stress that the structure (21) of the asymptotic vortex
configuration is dictated by purely symmetry and topological
arguments. It is independent of any specific choice of ansatz for the
matter or gauge fields. Inserting (21) in either (18) or (19) yields,
\begin{equation}
K = - \frac{\pi k n^{2}}{e^{2}}
\end{equation}
which is just the fractional spin reported earlier in [5,6]. It should
be realised that this is a noncanonical piece which, added to the
canonical angular momentum $J^{N}$, yields the physically relevant
(gauge invariant) angular momentum $J$. Since this piece is
independent of the origin of coordinates, it is interpreted as the
(fractional) spin (and not the orbital) part of the total angular
momentum.

It should be pointed out that (19) is the master equation which yields
the fractional spin of Chern-Simons vortices in any theory. This is
because the matter sector is completely absent in (19). Furthermore
$K$ is just a boundary term which reveals the topological origin of
fractional spin. In this sense this establishes a correspondence
with the approach of [1,8] where the argument for determining spin
depends on the topology associated with the relevant homotopy
class. An immediate fallout of (19) is that the result for the
spin should be unaffected by including the Maxwell term in the
original Lagrangian. This is because at large distances the low
derivative Chern-Simons term dominates over the higher derivative
Maxwell term. Consequently asymptotic effects (as, for instance,
given by (19)) are insensitive to the Maxwell term. This can also
be checked explicitly. Addition of the Maxwell piece modifies the
momenta (20) to,
\begin{equation}
\Pi_{i} = \frac{k}{2} \epsilon_{ij}A^{j} -
\frac{1}{e^{2}}(\partial_{\circ}A_{i} - \partial_{i}A_{\circ})
\end{equation}
The piece involving $A_{\circ}$ will not contribute to the
boundary term (19). This is because $A_{\circ}$ is a Lagrange
multiplier which can always be chosen so that the boundary term
vanishes. The asymptotic configuration (21) does not have any
explicit time dependence so that the piece
$\partial_{\circ}A_{i}$ also vanishes. Consequently the original
result for K (23) is reproduced. We note in passing that this
result for the Maxwell-Chern-Simons vortices was obtained earlier
[3,4] by using definite ansatz for matter and gauge fields in
either static or dynamic configurations.

Let us next consider the generalised Chern-Simons-Higgs model
introduced in [7]. It may be remarked that fractional spin of
these vortices has not been computed. Consequently this model
provides a virgin ground where our ideas can be tested and
compared with the usual approach [3-6]. The Lagrangian is defined
by [7],
\begin{equation}
{\cal L} = 2\sqrt{2}\epsilon^{\mu\nu\rho}[\eta^{4}A_{\rho} - 2i(\eta^{2}
-\frac{1}{2} \vert\phi\vert^{2})\phi (D_{\rho}\phi)^{*}]F_{\mu\nu} +
4(\eta^{2} - \vert\phi\vert^{2})^{2}\vert D_{\mu}\phi\vert^{2} -
V
\end{equation}
By requiring finiteness of energy it follows that $D_{i}\phi$
must vanish at spatial infinity, exactly as happens in the
previous model [6]. This observation is exploited to obtain the
asymptotic form of the momenta conjugate to $A_{i}$,
\begin{equation}
\pi^{i} = \frac{\partial {\cal L}}{\partial\dot{A}_{i}} =
4\sqrt{2}\epsilon^{ij}\eta^{4}A_{j}-8\sqrt{2}i\epsilon^{ij}(\eta^{2}
-\frac{1}{2}{\vert\phi\vert^{2}})(\phi(D_{j}\phi)^{*})
\end{equation}
as given by,
\begin{equation}
\pi^{i}
 \longrightarrow
4\sqrt{2}\epsilon^{ij}\eta^{4}A_{j}
\end{equation}
when $|x|\to\infty$. Using (21) and (27) in our master equation (19), 
the desired
expression for the fractional spin can be derived,
\begin{equation}
K = - 8\sqrt{2}\pi\eta^{4}n^{2}
\end{equation}

Let us next consider the usual method [3-6] of evaluating this spin.
The angular momentum given by Schwinger's [9] definition of the
energy momentum tensor,
\begin{equation}
J = \int d^{2}\vec{x} \epsilon^{ij}x_{i}\Theta_{\circ j}
\end{equation}
where,
\begin{equation}
\Theta_{\circ j} = \frac{\delta S}{\delta g^{\circ j}} =
4i(\eta^{2} - \vert\phi\vert^{2})^{2}A_{\circ}\phi(D_{j}\phi)^{*}
+ 2A_{j}\vert\phi\vert^{2}
\end{equation}
has to be computed in the static configuration. In this
configuration $A_{\circ}$ can be obtained from the equation of
motion [15],
\begin{equation}
\vert\phi\vert^{2}A_{\circ} =
-\frac{1}{\sqrt{2}(\eta^{2}-\vert\phi\vert^{2})}
\epsilon_{ij}[(\eta^{2}-\vert\phi\vert^{2})F_{ij}
-i\{D_{i}\phi(D_{j}\phi)^{*} - D_{j}\phi(D_{i}\phi)^{*}\}]
\end{equation}
For solutions with rotational symmetry the fields are defined by
the ansatz [6,7],
\begin{equation}
\phi(r,\theta) = \eta g(r)e^{in\theta}
\end{equation}
\begin{equation}
eA_{i}(r,\theta) = \epsilon_{ij}\frac{x^{j}}{\vert x
\vert^{2}}[a(r)-n]
\end{equation}
with the boundary conditions [6,7],
\begin{equation}
g(0) = 0, g(\infty) = 1
\end{equation}
\begin{equation}
a(0) = n, a(\infty) = 0
\end{equation}
Note that, asymptotically, (33) reduces to (21). Moreover the
ansatz (32, 33) is consistent with $D_{i}\phi
 \rightarrow 0$ as
$\vert x\vert\rightarrow\infty$,
which was stated earlier to follow from finiteness of energy.
Using (31) to (33) in (29, 30) one finds after some algebra,
\begin{equation}
J= 8\sqrt{2}\pi\eta^{4}\int^{\infty}_{\circ}dr\frac{d}{dr}[(1-g^{2})^{2}a^{2}]
\end{equation}
Finally, exploiting the boundary conditions (34, 35) yields,
\begin{equation}
J = - 8\sqrt{2}\pi\eta^{4}n^{2}
\end{equation}
which reproduces our earlier finding (28). The elegance and
algebraic  ease of our approach of computing the fractional spin
is clearly evident.

An important advantage of our method is its general
applicability. The conventional approach [3-6] of identifying
fractional spin with the angular momentum in a static
configuration is valid only if the vortex can be regarded as a
dyon, i.e., a flux-charge composite [1, 8]. For instance, it is
quite well known [8] that this analysis would lead to a vanishing
spin for the soliton in the O(3) nonlinear sigma model with the
Hopf term since $J$ is proportional to,
\begin{equation}
J \sim \int d^{2}\vec{x} \epsilon_{ij} x_{i} \theta_{\circ j} = \int
d^{2}\vec{x} \epsilon_{ij}x_{i}(\dot{n}^{a}\partial_{j}n^{a})
\end{equation}
which vanishes in the static configuration $(\dot{n}^{a} = 0)$ where
$n^{a}$ are the basic fields. On the contrary it is well
established [1, 2] that this soliton does have a fractional spin.
If, however, we proceed to compute the fractional spin as $J -
J^{N}$, the correct result can be reproduced [14]. The fact that
the fractional spin of a vortex (that may be interpreted as a
dyon) comes out correctly as the (Schwinger) angular momentum $J$
in the static configuration is related to the finding (to be
discussed below) that the (Noether) angular momentum $J^{N}$ in
such a configuration vanishes. In that case the usual analysis
may be identified with our approach of computing $(J - J^{N})$ in
the particular case of a static configuration. We now explicitly
show how $J^{N}$ vanishes for the model (1). Inserting the
expressions for $\phi$ and $A_{i}$ in the static configuration
using (32, 33) in (17), one finds,
\begin{equation}
J^{N} = -\frac{kn}{e}\int d^{2}\vec{x}
\epsilon_{ij}x_{i}\partial_{j}\Theta\frac{1}{er}\frac{da(r)}{dr}
\end{equation}
which can be expressed, exploiting symmetry arguments, as a total
boundary,
\begin{eqnarray}
J^{N} &= &-\frac{kn}{e} \int d^{2}\vec{x}
\partial_{j}[\epsilon_{ij}x_{i} \frac{a'(r)}{er}\theta] \nonumber
\\
      &= &-\frac{kn}{e} \int
      dx_{1}dx_{2}\{x_{1}\partial_{2}(\frac{a'(r)}{er}\theta) -
      x_{2}\partial_{1}(\frac{a'(r)}{er}\theta)\}
\end{eqnarray}
Taking the value of $\frac{a'(r)}{r}$ from ref.[6],
\begin{equation}
\frac{a'(r)}{r} \sim  g^{2} (g^{2} - 1)
\end{equation}
which vanishes asymptotically since $g(\infty) = 1$ (34). Hence
$J^{N}$ also vanishes.

Let us finally extend our analysis to analyse fractional spin in
non-abelian vortex configurations. Proceeding as in the abelian
models, one can show that the fractional spin is given by the
non-abelian analogue of (19),
\begin{equation}
K = - \int d^{2}\vec{x} \partial^{i}[\pi_{ia} \epsilon^{jk}x_{j}
A_{k}^{a}]
\end{equation}
where `a' denotes the group index. The asymptotic form of the
nonabelian vortex configuration [16] is structurally identical to
(21),
\begin{equation}
A^{a}_{i} = - \epsilon_{ij} \frac{x^{j}}{\vert x \vert^{2}}
\frac{Q^{a}}{2\pi k}
\end{equation}
where,
\begin{equation}
Q^{a} = \int d^{2}\vec{x}J_{\circ}^{a}
\end{equation}
with $J^{a}_{\mu}$ being the nonabelian current,
\begin{equation}
J^{a}_{\mu} = \phi^{*} T^{a}(D_{\mu}\phi) - (D_{\mu}\phi)^{*}
T^{a}\phi
\end{equation}
which enters the Euler-Lagrange equation as,
\begin{equation}
k\epsilon^{\mu\beta\sigma}(\partial_{\beta}A^{a}_{\sigma} +
\frac{1}{2} f^{abc} A^{b}_{\beta} A^{c}_{\sigma}) = J^{\mu a}
\end{equation}
The normalisation in (43) is determined [16] from the fact that it
should obey the charge-flux identity following from the time-component
of (46). Inserting (43) in (42) yields the expression,
\begin{equation}
K = - \frac{\pi k Q^{a} Q_{a}}{(2\pi k)^{2}}
\end{equation}
Till now the analysis is completely general because the gauge group
has not yet been specified. If we consider this group to be SU(2),
which is the simplest example admitting nonabelian vortices, it is
possible to express $Q^{a} Q_{a}$ in (47) in terms of the vortex
number $n$. Requiring the finiteness of energy demands that
$(D_{i}\phi)_{m}$ should vanish asymptotically, in analogy with the
abelian case. Using the ansatz (32) and (43) this implies,
\begin{equation}
[(q_{a}T^{a} + in)\eta]_{m} = 0;  q_{a} = -\frac{Q_{a}}{2\pi k}
\end{equation}
For nontrivial solutions of $\eta$ this yields the condition,
\begin{equation}
\det\vert(q_{a}T^{a} + in)\vert = 0
\end{equation}
For SU(2), the $T^{a}$ are just the Pauli matrices. Using their
standard representation, one obtains from (49),
\begin{equation}
q_{a}q^{a} = n^{2}
\end{equation}
Substituting in (47) yields,
\begin{equation}
K = - \pi k n^{2}
\end{equation}
which is structurally identical to the abelian result (23). There is a
difference of $e^{2}$ because the coupling constant was absorbed in
the definition of the gauge potential (43). For the particular case of
$n = 1$, where detailed vortex configurations have been worked out
[4], the result agrees with (51).

To conclude, we have made a detailed investigation of one particularly
interesting characteristic of Chern-Simons vortices; namely, the
(fractional) spin carried by these vortices. A master formula (19) was
obtained which was used to compute this spin in different Chern-Simons
theories. The crucial role played by the Gauss constraint was
emphasised. The basic idea was to modify the gauge invariant
(Schwinger's) energy momentum tensor by adding a term proportional to
the Gauss constraint such that the covariant transformation properties
of the basic fields were preserved. Next, the physical (gauge
invariant) angular momentum was obtained from this tensor. The total
canonical (orbital plus spin) angular momentum was abstracted from the
physical angular momentum by subtracting the Noether angular momentum.
The remainder was a total boundary and, being independent of the
origin of co-ordinates, was interpreted as the (fractional) spin of
the Chern-Simons vortex. Since this spin is a boundary effect, its
topological origin is revealed. Explicit computation just required the
knowledge of the asymptotic form of the gauge potential in a
rotationally symmetric configuration. The final answer is, therefore,
structurally identical in any Chern-Simons theory. Furthermore,
addition of the Maxwell term does not affect our conclusions since, in
the asymptotic limit, the Chern-Simons term dominates. A comparison of
our analysis with the usual methods [3-6] of computing fractional spin
was also done. Finally it was shown how nonabelian vortex
configurations could be analysed within the present scheme.

\end{document}